\documentclass[a4paper,11pt]{article}
\pdfoutput=1 

\usepackage{jheppub} 

\usepackage{slashbox}
\usepackage[T1]{fontenc} 
\usepackage{slashed}

\preprint{arXiv:1605.xxxxx [math-ph]}

\title{The graded product of real spectral triples}
  
\author{Shane Farnsworth}
\affiliation{$^2$Max Planck Institute for Gravitational Physics,\\
Am M$\ddot{\text{u}}$hlenberg 1, 14476 Golm, Germany, EU}  

\abstract{Forming the product  of two geometric spaces   is one of the most basic operations in geometry, but in the spectral-triple formulation of non-commutative geometry, the standard prescription for taking the product of two real spectral triples is problematic: among other drawbacks, it is non-commutative, non-associative, does not transform properly under unitaries, and often fails to define a proper spectral triple.  In this paper, we explain that these various problems result from using the {\it ungraded} tensor product; by switching to the {\it graded} tensor product, we obtain a new prescription where all of the earlier problems are neatly resolved: in particular, the new product is commutative, associative, transforms correctly under unitaries, and always forms a well defined spectral triple.}

\begin{document}
\maketitle


\section{Introduction}
\label{Sec_Intro}

Forming the product between two geometric spaces is a basic operation in geometry. In non-commutative geometry (NCG) not only do product geometries provide a rich set of example spaces for mathematicians to explore, but they are also of great physical interest, because  they arise in the description of gauge theories (including the standard model of particle physics, and its extensions) coupled to Einstein gravity. Unfortunately, in the spectral triple formulation of NCG, the traditional prescription for taking the product of geometric spaces has problems. In particular, although it should be expected that the product operation be commutative and associative, and to transform naturally under unitaries, it does not; and given two geometries $T_1$ and $T_2$ with well-defined `KO-dimensions' $d_1$ and $d_2$ respectively, it should be expected that their product $T_{1,2} = T_1\times T_2$ also has a well defined KO dimension $d_{i,j }=d_1+d_2$ (mod 8), but in general it does not. In this paper, we point out that these difficulties (and others) ultimately result from the fact that the traditional prescription incorrectly uses the {\it ungraded} tensor product to describe the product between graded spaces. We show that by switching to the {\it graded} tensor product, all of these issues are neatly resolved.  

The paper is organised as follows: In Sections~\ref{Sec_Trad} and~\ref{Sec_Grad_Prod} we cover review material. In particular  we start in  Subsection~\ref{Sec_Trad_KO_Dim} by briefly reviewing the idea of KO-dimension, after which in Subsection~\ref{Sec_Trad_NCG} we  review the traditional prescription for taking the product between two or more real NCGs and describe what goes wrong in general. In Section~\ref{Sec_Grad_Prod} we review graded tensor products as they are defined for star differential graded algebras ($*$-DGAs). The material from Section~\ref{Sec_Prod_New} onwards is new. In Sections~\eqref{Sec_Prod_ee},~\eqref{Sec_Prod_eo}, and~\eqref{Sec_Prod_oo} we use the graded tensor product developed for $*$-DGAs to redefine the product between NCGs. Then in Subsection~\ref{Prod_Mneminic} we provide a useful mnemonic for constructing the full KO-dimension table. In Subsection~\ref{Sec_Conclusion} we  briefly detail how our new prescription builds upon the previous approaches introduced in~\cite{2011IJGMM..08.1833D,Cacic:2012qj, Vanhecke99,Sitarz}.

\section{The traditional product prescription (and its shortcomings)}
\label{Sec_Trad}

The goal of this section is two fold: (i) We begin in Subsection~\ref{Sec_Trad_KO_Dim} by reviewing the idea of KO-dimension, which is the notion of dimension we will be making use of in this paper. (ii) In Subsection~\ref{Sec_Trad_NCG} we review the traditional prescription for taking the product between two real spectral triples in NCG. We show what goes wrong, and briefly discuss some of the previous proposed solutions.

\subsection{KO-dimension}
\label{Sec_Trad_KO_Dim}
There are several equally good ways of defining the dimension of an ordinary Riemannian geometry. The most familiar definition is given in terms of the number of linearly independent basis elements in the tangent space of a smooth manifold. Alternatively one could instead make use of Weyl's law, which relates the asymptotic growth of the eigenvalues of the Laplace operator on a manifold to the metric dimension of the underlying space~\cite{Marcollibookdraft}. A less familiar notion is so called `KO-dimension', which exists for Riemannian spin geometries and more generally for `real' NCGs. KO-dimension can be understood in the following functional sense: Consider a familiar four dimensional Riemannian  spin geometry equipped with the flat Dirac operator $D = -i\gamma^\mu\partial_\mu$, the Dirac gamma five matrix $\gamma= \gamma^0\gamma^1\gamma^2\gamma^3$, and the charge conjugation operator $J = \gamma^0\gamma^2\circ cc$, where we are using the basis of hermitian gamma matrices $\gamma^a$ given in~\cite[$\S3.4$]{Schucker:2001aa}.  If one checks, then what one finds is that the operators $\{D,J,\gamma\}$ satisfy the following  conditions:
\begin{align}
J^2 = \epsilon\mathbb{I},\hspace{1.5cm}JD = \epsilon'DJ,\hspace{1.5cm}J\gamma = \epsilon''\gamma J,\label{Eq_Real_Structure}
\end{align}
where  $\{\epsilon,\epsilon',\epsilon''\}=\{-1,+1,+1\}$. An analogous treatment can be performed in any metric dimension~\cite[\S B]{Polchbook}, however in general the signs $\{\epsilon,\epsilon',\epsilon''\}$ will depend on the dimension mod $8$ of the underlying manifold\footnote{As explained in~\cite{Barrett:2006qq,Connes:2006qv}, `KO-dimension' is a misnomer, and really only corresponds to metric dimension for commutative Riemannian geometries. If for example we had instead considered the familiar $4D$ Lorentzian spin geometry with  Dirac operator $D = -i\gamma^\mu\partial_\mu$, gamma five matrix $\gamma= i\gamma^0\gamma^1\gamma^2\gamma^3$, and a charge conjugation operator $J_U = \gamma^2\circ cc$, then we would have found signs $\{\epsilon,\epsilon',\epsilon''\}$ corresponding to the `KO-signature' $3-1=2$ case.}. Said another way, the signs $\{\epsilon,\epsilon',\epsilon''\}$ \textit{define} the `KO-dimension' of a spin geometry, and this idea continues to make sense for real NCGs. The notion of KO-dimension has many deep connections with Clifford algebras, Bott-periodicity, homology, etc (see e.g.~\cite{GVF2000,Polchbook,2011IJGMM..08.1833D,
ATIYAH1969,Connes:1996gi,ConnesMarcolliBook,
ATIYAH19643}), but the functional definition outlined here is all that will be necessary for understanding the rest of the paper.  In table~\ref{Tab_KO_Connes} we collect the various signs corresponding to each KO-dimension as they are usually presented in the NCG literature.

\begin{table}[h]
\begin{center}
  \begin{tabular}{| c | c | c | c | c | c | c | c | c  |}
    \hline
     & $0$ & $1$ & $2$ & $3$ & $4$ & $5$ & $6$ & $7$ \\ \hline\hline
      $\epsilon$    & $+1$ & $+1$ & $-1$ & $-1$ & $-1$ & $-1$ & $+1$ & $+1$ \\ 
      \hline
            $\epsilon'$   & $+1$ & $-1$& $+1$ & $+1$& $+1$ & $-1$& $+1$ & $+1$ \\ 
            \hline
                  $\epsilon''$   & $+1$ &  & $-1$ &  & $+1$ &  & $-1$ &    \\
                  \hline
  \end{tabular}
\end{center}
\caption{\label{Tab_KO_Connes}\textit{Mod 8 KO-dimension table} as it is traditionally presented in the NCG literature (see e.g.~\cite{Connes:2006qv,Chamseddine:2006ep,
ConnesMarcolliBook,Barrett:2006qq,
vandenDungen:2012ky,Connes:1996gi,
GVF2000}). With this presentation no obvious patterns emerge in the  signs.} 
\end{table}

\subsection{Product non-commutative geometries}
\label{Sec_Trad_NCG}

 NCG is a generalization of Riemannian geometry which (amongst other applications) provides an elegant framework for describing gauge theories coupled to gravity. In this capacity, it's main physical interest is in constraining the allowed extensions of the standard model of particle physics~\cite{ 
Boyle:2014wba,Farnsworth:2014vva,Boyle:2016cjt,Brouder:2015qoa,Stephan:2009te,Stephan:2013eda,
Chamseddine:2007hz,Chamseddine:2007ia,Chamseddine:2006ep,Connes:1996gi,
Krajewski:1996se,Chamseddine:2013rta,Devastato:2013oqa}.
The basic idea of NCG is to replace the familiar manifold and metric data $\{M,g\}$ of Riemannian geometry with a `spectral triple' of data $\{A,H,D\}$, which consist of a `coordinate' algebra $A$ that provides topological information, a Dirac operator $D$ which provides metric information, and a Hilbert space $H$ that provides a place for $A$ and $D$ to interact. A spectral triple is said to be `real' and `even' if it is also equipped with an anti-unitary real structure operator $J$~\cite{Connes:1995tu} and a $\mathbb{Z}_2$ grading operator $\gamma$ on $H$ respectively. We will call a spectral triple which is not equipped with a non-trivial grading operator `odd'. The benefit of this `spectral' approach to geometry is that it continues to make sense even when the input algebra $A$ is non-commutative, hence the name `non-commutative geometry'. For a  review see for example~\cite{vandenDungen:2012ky,
ConnesMarcolliBook,Schucker:2001aa,
vanSuijlekomBook,Connes:1996gi,LandiBook,ConnesBook}.

To build a sensible NCG, the data $\{A,H,D,J,\gamma\}$ should not be selected arbitrarily, but instead must satisfy a number of geometric conditions and axioms (which generalize the conditions satisfied by commutative Riemannian geometries, see e.g.~\cite{Connes:1996gi,Chamseddine:2006ep,
ConnesMarcolliBook} for details). In particular, a `real' NCG must have a well defined KO-dimension, which in practice means that  the operators $\{D,J,\gamma\}$ must satisfy the `real structure' conditions outlined in Eqs.~\eqref{Eq_Real_Structure} for an appropriate set of signs $\{\epsilon,\epsilon',\epsilon''\}$. A useful trick for finding new and interesting geometric spaces which satisfy the NCG axioms is to build product geometries from spaces which are already known to satisfy the NCG axioms. Unfortunately, as we will discuss now, the traditional prescription for taking the product between two or more real spectral triples does not always result in a product space with a well defined KO-dimension.

In the traditional prescription, a product NCG is defined as follows: Given two real spectral triples  $T_i = \{A_i,H_i,D_i,J_i,\gamma_i\}$ and 
$T_j = \{A_j,H_j,D_j,J_j(,\gamma_j)\}$, the first of which is necessarily even, their product $T_{i}\times T_j$ is defined by $T_{i, j} = \{A_{i, j},H_{i, j}, D_{i, j},J_{i, j}(,  \gamma_{i, j})\}$ where~\cite{Connes:1995tu,Connes:1996gi,ConnesMarcolliBook,2011IJGMM..08.1833D, Cacic:2012qj}:
\begin{subequations}
\begin{align}
A_{i, j} = A_i\widehat{\otimes} A_j,\hspace{1cm}
H_{i, j} = H_i\widehat{\otimes} H_j,\nonumber\hspace{0.0cm}\\
D_{i, j} = D_i\widehat{\otimes}\mathbb{I}_j + \gamma_i\widehat{\otimes} D_j,\hspace{1cm}\label{Eq_Prod_Orig_1}\\
\gamma_{i, j} = \gamma_i\widehat{\otimes} \gamma_j,\hspace{1.4cm} J_{i, j} = J_i\widehat{\otimes} J_j,\nonumber
\end{align}
and where $\widehat{\otimes}$ is the usual tensor product (see e.g.~\cite[\S 4]{2011IJGMM..08.1833D}), and the $\mathbb{Z}_2$ grading operator $\gamma_{i,j}$ is only defined if both $T_i$ and $T_j$ are even.\label{Eq_Prod_Orig}

The product given in Eqs.~\eqref{Eq_Prod_Orig_1} does not always form a well defined spectral triple satisfying the real structure conditions of  Eqs.~\eqref{Eq_Real_Structure}. In fact for the signs given in Table~\ref{Tab_KO_Connes}, this product only makes sense if the first spectral triple $T_i$ is of KO-dimension $0$ or $4$ (mod 8): When $T_i$ is of KO-dimension $2$ or $6$ (i.e. when $\epsilon_i'' = -1$) then the product geometry $T_{i,j}$ fails to satisfy the real structure condition $J_{i,j}D_{i,j} = \epsilon_{i,j}'D_{i,j}J_{i,j}$ of Eqs.~\eqref{Eq_Real_Structure}, and when the first spectral triple $T_i$ is of odd KO-dimension the product Dirac operator $D_{i,j}$ is not defined at all (because an odd $T_i$ will not be equipped with a non-trivial grading operator $\gamma_i$).
Worse still, the definitions given in Eqs.~\eqref{Eq_Prod_Orig} are inherently non-symmetric in the sense that even when a product geometry $T_{i,j}$ is well defined, $T_{j,i}$ is not necessarily. 

A partial solution to the above mentioned problems is  obtained if one makes two important observations: (i) The first observation, which was emphasised in~\cite{Vanhecke99,2011IJGMM..08.1833D}, is that for even spectral triples there is a second equally good choice for the product Dirac operator:
\begin{align}
\widetilde{D}_{i, j} &= D_i\widehat{\otimes}\gamma_j + \mathbb{I}_i\widehat{\otimes} D_j.
\end{align}
\end{subequations}
The two choices of Dirac operator given in Eqs.~\eqref{Eq_Prod_Orig} are unitarily equivalent, $\widetilde{D}_{i,j} = UD_{i,j}U^*$, with the unitary operator $U$ given by~\cite{Vanhecke99}:
\begin{align}
U = \tfrac{1}{2}(\mathbb{I}_i\widehat{\otimes}\mathbb{I}_j + \gamma_i\widehat{\otimes}\mathbb{I}_j + \mathbb{I}_i\widehat{\otimes}\gamma_j - \gamma_i\widehat{\otimes}\gamma_j) \label{Eq_transformU}.
\end{align}
(ii) The second observation which was emphasised in~\cite{2011IJGMM..08.1833D} is that Table~\ref{Tab_KO_Connes} should be extened to include 12 instead of 8 possible KO-dimension signs. This is because in each even KO-dimension there are two equally good ways of defining the real structure operator: If $J_U$ is a real structure operator with KO-dimension signs $\{\epsilon_U,\epsilon_U',\epsilon_U''\}$, then the composition $J_L =\gamma J_U$ is also an anti-unitary operator satisfying the real structure conditions given in~\eqref{Eq_Real_Structure} with  signs $\{\epsilon_L''\epsilon_L,-\epsilon_L',\epsilon_L''\}$. The `U' and `L' subscripts stand for `upper' and `lower' respectively - the reason for our naming convention will become apparent in Section~\ref{Sec_Prod_New}. For odd spectral triples the grading operator is trivial $\gamma \propto \mathbb{I}$, and the upper and lower sign choices $\{\epsilon,\epsilon'\}$ are degenerate.
\begin{table}[h]
\begin{center}
  \begin{tabular}{| c | c | c | c | c || c | c | c | c  ||c|c|c|c|}
    \hline
     & $0$ & $2$ & $4$ & $6$ & $0$ & $2$ & $4$ & $6$ & $1$ & $3$ & $5$ & $7$\\ \hline\hline
      $\epsilon$    & $+1$ & $+1$ & $-1$ & $-1$ & $+1$ & $-1$ & $-1$ & $+1$  & $+1$ & $-1$ & $-1$ & $+1$ \\ 
      \hline
            $\epsilon'$   & $-1$ & $-1$& $-1$ & $-1$& $+1$ & $+1$& $+1$ & $+1$ & $-1$ & $+1$& $-1$ & $+1$\\ 
            \hline
                  $\epsilon''$   & $+1$ & $-1$  & $+1$ & $-1$ & $+1$ & $-1$  & $+1$ &$-1$  &  & &  &   \\
                  \hline
                    & $L$ & $U$  & $L$  & $U$  & $U$ & $L$  & $U$ &$L$  &  & &  &  \\\hline
  \end{tabular}
\end{center}
\caption{\label{Tab_KO_Dabrow}\textit{Extended Mod 8 KO-dimension table} as presented in~\cite{2011IJGMM..08.1833D}, with even KO-dimension signs grouped according to their $\epsilon'$ sign. `Even' KO-dimension signs corresponding to our `upper' (`lower') naming convention are marked with a `U' (`L').} 
\end{table}

When taken together, these two observations  extend the applicability of the product defined in Eqs.~\eqref{Eq_Prod_Orig} significantly~\cite{Vanhecke99,2011IJGMM..08.1833D}. For example, if the product between a certain pair of even spectral triples $T_i$ and $T_j$ is not well defined, then one can always find a well defined product triple $T_{i,j}$ by first replacing either the real structure operator $J_i$ with $\gamma_i J_i$, or by replacing  $J_j$ with $\gamma_j J_j$ (i.e. if the product between triples $T_i$ and $T_j$ is not well defined, then replacing $T_i = \{A_i,H_i.D_i,J_i,\gamma_i\}$ with $\widetilde{T}_i = \{A_i,H_i.D_i,\gamma_iJ_i,\gamma_i\}$ in the product will always result in a well defined geometry $T_{i,j}$). Similarly, products which are poorly defined when using the Dirac operator $D_{i,j}$ may make sense if instead the unitariliy equivalent choice of Dirac operator $\widetilde{D}_{i,j}$ is used. What is more,  the definitions given in Eqs.~\eqref{Eq_Prod_Orig} have been extended to include the odd-odd cases in~\cite{2011IJGMM..08.1833D,Sitarz,Cacic:2012qj}.  Despite these improvements, the product as defined in Eqs.~\eqref{Eq_Prod_Orig} remains problematic:
\begin{itemize}
\item \textbf{Undefined products:} For even spectral triples there are two equally good choices for the real structure operator $\{J,\gamma J\}$. Therefore when forming the product of any two real, even spectral triples there are four possible combinations for the product real structure operator (i.e. $J_i\widehat{\otimes} J_j,\gamma_iJ_i\widehat{\otimes} J_j,J_i\widehat{\otimes} \gamma_jJ_j,$ or $\gamma_iJ_i\widehat{\otimes} \gamma_jJ_j$), while only two of these four possibilities may correspond to a well defined product geometry. To understand what goes wrong for two of the four choices it is useful to examine the KO-dimension signs $\{\epsilon_{i, j},\epsilon_{i, j}',\epsilon_{i, j}''\}$ corresponding to a product space $T_{i,j} = T_i\times T_j$. For the definitions given in Eqs.~\eqref{Eq_Prod_Orig} these are given by:
\begin{subequations}
\begin{equation}
\epsilon_{i, j} = \epsilon_i\epsilon_j,\hspace{1.5cm}\epsilon_{i, j}' = \epsilon_i' = \epsilon_i''\epsilon_j',\hspace{1.5cm} \epsilon_{i, j}'' = \epsilon_i''\epsilon_{j}'', 
\end{equation}
or
\begin{equation}
\widetilde{\epsilon}_{i, j} = \epsilon_i\epsilon_j,\hspace{1.5cm}\widetilde{\epsilon}_{i, j}' = \epsilon_i'\epsilon_j'' = \epsilon_j',\hspace{1.5cm} \widetilde{\epsilon}_{i, j}'' = \epsilon_i''\epsilon_{j}''.
\end{equation}\label{Eq_symprod}\end{subequations}
where the product signs with `tildes' $\{\widetilde{\epsilon}_{i,j},\widetilde{\epsilon}'_{i,j},\widetilde{\epsilon}''_{i,j}\}$ correspond to the choice of Dirac operator $\widetilde{D}_{i,j}$, while those without tildes $\{{\epsilon}_{i,j},{\epsilon}'_{i,j},{\epsilon}''_{i,j}\}$ correspond to the choice $D_{i,j}$. It is clear from Eqs.~\eqref{Eq_symprod} what must go wrong: For certain real structure combitations it is not possible to satisfy $\epsilon_{i, j}' = \epsilon_i' = \epsilon_i''\epsilon_j'$ and/or $\widetilde{\epsilon}_{i, j}' = \epsilon_i'\epsilon_j'' = \epsilon_j'$. In tables 2-5 of~\cite{2011IJGMM..08.1833D}, and 2-5 of~\cite{Vanhecke99} the authors give a full listing of which product geometries have a well defined KO-dimension, along with those which do not.

\item \textbf{Transformation under unitaries:} Despite the two Dirac operators $D_{i, j}$ and $\widetilde{D}_{i, j}$ being unitarily equivalent, it \textit{does} matter which one is used when taking the product of two even spaces~\cite{2011IJGMM..08.1833D}. While some products are always well defined regardless of which Dirac operator is selected, others depend on the choice between $D_{i, j}$ and $\widetilde{D}_{i, j}$, while other products  are never well defined. In addition, product triples as defined in Eqs.~\eqref{Eq_Prod_Orig}  are not stable under the unitary transformation  of the Dirac operator given in Eq.~\eqref{Eq_transformU}, in the sense that while the product algebra $A_{i, j}$ and grading $\gamma_{i, j}$  are invariant under conjugation by $U$, the real structure operator $J_{i, j}$  is not. It transforms along with the Dirac operator. 

\item \textbf{Commutativity and Associativity:} The product defined in Eqs.~\eqref{Eq_Prod_Orig} is non-commutative in the sense that while $T_{i,j}$ may be well defined, $T_{j,i}$ is not necessarily. Perhaps more troubling however is that  the product is not associative, in the sense that while a product $(T_i\times_D T_j)\times_{\widetilde{D}}T_k$ may be well defined, the product $T_i\times_D (T_j\times_{\widetilde{D}}T_k)$ is not necessarily (where the $D$ and $\widetilde{D}$ subscripts indicate which choice of Dirac operator is being used for the product).

\item \textbf{Obscure grading factors:} The two product Dirac operators defined in Eqs.~\eqref{Eq_Prod_Orig} include  grading factors. These factors are introduced to ensure that the total Dirac operator squares to $D_{i,j}^2 = D_i^2\widehat{\otimes}\mathbb{I}_j + \mathbb{I}_i\widehat{\otimes} D_j^2$, which implies that the dimensions add $d_{i,j} = d_i+d_j$~\cite{Vanhecke99}. Grading factors also appear when translating between `upper' and `lower' real structure operators $J_L=\gamma J_U$. The distinction between `upper' and `lower' spectral triples and between $D_{i,j}$ and $\widetilde{D}_{i,j}$ \textit{does} seem to matter, and so it would be good to understand what is it that governs the appearance of the various grading factors in well defined product geometries.

\item \textbf{Obscure KO-dimension signs:} The product as defined in Eqs.~\eqref{Eq_Prod_Orig} together with the KO-dimension table as presented in Table~\ref{Tab_KO_Dabrow}, provides little hint as to why certain products work, and why others fail. There is no obvious pattern behind the various KO-dimension signs, and no good reason for distinguishing those even signs for which $\epsilon'=+1$ from those satisfying $\epsilon'=-1$ as is done in the literature (see e.g. e.g.~\cite{Connes:2006qv,Chamseddine:2006ep,
ConnesMarcolliBook,Barrett:2006qq,
vandenDungen:2012ky,Connes:1996gi,
GVF2000}).

\end{itemize}

A number of solutions to the above mentioned problems have already been proposed. In particular the authors in~\cite{Vanhecke99,Sitarz,Cacic:2012qj} provide new definitions for the product real structure operator $J_{i,j}$, each of which includes various clever insertions of grading factors $\gamma_i$ and $\gamma_j$, which depend explicitly on the KO-dimensions of the two spectral triples being multiplied. While it is always possible to form well defined products in this way, the definitions already proposed offer no real explanation for the various obscure grading factors which are forced to appear. They also either depend on lookup tables, or unnaturally distinguish those KO-dimension signs for which $\epsilon'$ is positive. Stability of the various definitions under the unitary transformation given in Eq.~\eqref{Eq_transformU} has also not been discussed. In Section~\ref{Sec_Prod_New} we will show that a much more natural definition for the product between spectral triples is given in terms of the graded tensor product. The new definitions we provide are simple, and neatly resolve \textit{all} of the various problems and questions which arise for the product defined in Eqs.~\eqref{Eq_Prod_Orig}.

\section{Graded tensor products}
\label{Sec_Grad_Prod}

The purpose of this section is to provide a brief review of $*$-DGAs, as well as to review graded tensor products as they are defined for $*$-DGAs. For a more complete account see the second section and the appendix of~\cite{Boyle:2016cjt}.

\subsection{Differential graded star algebras}
\label{Sec_Prelim_DGA}

A $\mathbb{Z}$ graded vector space $H$ (over a field $\mathbb{F}$), is a vector space which decomposes into the direct sum of vector spaces $H_i$ (each defined over the field $\mathbb{F}$):
\begin{align}
H = \bigoplus_{i\in\mathbb{Z}}H_i.
\end{align}
Any element $h\in H_i$ is said to be of `degree' or `order' $|h| = i\in\mathbb{Z}$.

A graded algebra $A$ over the field $\mathbb{F}$, is defined to be a graded vector space over $\mathbb{F}$ which is equipped with a bi-linear product over  $\mathbb{F}$, $A\times A\rightarrow A$, which respects the grading on $A$ in the sense: $|aa'|  = |a| + |a'|$ for $a,a'\in A$. 
 
A graded algebra $A$ is said to be involutive if it is equipped with an anti-linear operator $*:A\rightarrow A$ which satisfies:
\begin{subequations}
\begin{align}
(a^*)^* &= a,\\
(aa')^* &= (-1)^{|a|.|a'|}{a'}^*a^*,\label{involcond}
\end{align}\label{involprop}\end{subequations}
for $a,a'\in A$.\footnote{Note that our choice of sign convention here corresponds to `convention 2' as outlined in~\cite{Boyle:2016cjt}.} 
A graded algebra is said to be differential if it is equipped with a linear first order differential operator $d:A\rightarrow A$, which satisfies:
\begin{subequations}
\begin{align}
d^2 &= 0,\\
d[aa'] &= d[a]a' + (-1)^{|a|}ad[a'] 
\end{align}\label{dcond}\end{subequations}
for $a,a'\in A$. An algebra $A$  is said to be a $*$-DGA if it is equipped with an involution $*$ and a differential $d$ satisfying Eqs.~\eqref{involprop} and~\eqref{dcond} respectively, along with the condition
\begin{align}
d[a^*] = \pm  d[a]^*,\label{dconnew}
\end{align}
for $a\in A$.\footnote{For a natural generalization of condition~\eqref{dconnew} see~\cite{Boyle:2016cjt}.}

\subsection{Graded tensor products}
\label{Prelim_grad_tens}

The action of linear operators on graded vector spaces can be defined in the same way as is done for spaces which are ungraded. In particular, a linear operator $\mathcal{O}$ on a graded vector space $H$ is a map from $H$ to itself satisfying:
\begin{subequations}
\begin{align}
\mathcal{O}(\alpha_1 h_1 + \alpha_2 h_2) &= \alpha_1\mathcal{O} h_1 + \alpha_2\mathcal{O} h_2 ,
\end{align}
\end{subequations}
where $h_1,h_2\in H$, and $\alpha_1,\alpha_2\in \mathbb{F}$. An operator $\mathcal{O}$ is said to be of `degree' or `order' $|\mathcal{O}| =j\in\mathbb{Z}$ if it maps elements of $H_i$ into elements of $H_{i+j}$, i.e. $\mathcal{O}:H_i\rightarrow H_{i+j}$. Notice that any element $a\in A_j$ of a graded algebra $A$ (as defined above in Subsection~\ref{Sec_Prelim_DGA}) can be thought of as  an operator of degree $j$ on $A$, i.e. $a:A_i\rightarrow A_{i+j}$. 

Given two graded vector spaces $H'$ and $H''$ over the field $\mathbb{F}$ and graded linear operators $\mathcal{O}':H'\rightarrow H'$ and $\mathcal{O}'':H''\rightarrow H''$ respectively, their graded tensor product is defined as follows: the product vector space $H$ is the tensor product of the vector spaces $H'$ and $H''$, where the degree of an element $h'\otimes h''\in H'\otimes H''$ is defined to be $|h'\otimes h''| \equiv |h'| + |h''|$. The product operator $\mathcal{O}'\otimes \mathcal{O}'':H'\otimes H''\rightarrow H'\otimes H''$ is defined to be of order  $|\mathcal{O}'\otimes \mathcal{O}''| = |\mathcal{O}'| + |\mathcal{O}''|$, while its action on $H$ is defined such that:
\begin{subequations}
\begin{align}
(\mathcal{O}'\otimes \mathcal{O}'')(h'\otimes h'') \equiv (-1)^{|\mathcal{O}''||h'|}(\mathcal{O}'h'\otimes \mathcal{O}''h''),\label{convention1}
\end{align}
or alternatively:
\begin{align}
(\mathcal{O}'\otimes \mathcal{O}'')(h'\otimes h'') \equiv (-1)^{|\mathcal{O'}||h''|}(\mathcal{O'}h'\otimes \mathcal{O}''h''),\label{convention2}
\end{align}\label{gradedaction}\end{subequations}
for $h\in H$, $h'\in H'$. The choice between the `Kozul' signs given in Eqs.~\eqref{gradedaction} is purely conventional, but will be of consequence when we later define the graded product between NCGs\footnote{The ungraded tensor product is defined with no `Kozul' sign, i.e. $(\mathcal{O}'\widehat{\otimes} \mathcal{O}'')(h'\widehat{\otimes} h'') = (\mathcal{O'}h'\widehat{\otimes} \mathcal{O}''h'')$, see e.g.~\cite[\S 4]{2011IJGMM..08.1833D}.}. It is easy to show that the graded tensor product is associative.

The definitions given in~\eqref{gradedaction} are all that is needed to construct the graded tensor product of two $*$-DGAs. Given two graded algebras $A'$ and $A''$, the order of an element $a'\otimes a''\in A'\otimes A''$ is defined to be $|a'\otimes a''| = |a'| + |a''|$. Multiplication between any two elements $a_1'\otimes a_1''$ and $a_2'\otimes a_2''$ in $A'\otimes A''$ is defined following~\eqref{gradedaction} to be:
\begin{subequations}
\begin{align}
(a_1'\otimes a_1'')(a_2'\otimes a_2'')\equiv (-1)^{|a_1''||a_2'|}(a_1' a_2'\otimes a_1'' a_2''),  \label{convA1}
\end{align}
or alternatively:
\begin{align}
(a_1'\otimes a_1'')(a_2'\otimes a_2'')\equiv (-1)^{|a_1'||a_2''|}(a_1' a_2'\otimes a_1'' a_2''),  
\end{align}
depending on the `Kozul' sign convention chosen. If $A'$, and $A''$ are equipped with star operations $*'$, and $*''$ respectively, then the star operation on the product algebra $A = A'\otimes A''$ is defined to be:
\begin{align}
* =*'\otimes *''.\label{eq_invol}
\end{align}
If $A'$, and $A''$ are equipped with differential operators $d'$ and $d''$ respectively, then the differential on the product algebra $A = A'\otimes A''$ is defined to be:
\begin{align}
d=d'\otimes \mathbb{I}'' + \mathbb{I}'\otimes d''.\label{eq_dirac}
\end{align}\label{star-d}\end{subequations}
The graded tensor product as given in Eqs.~\eqref{gradedaction} is defined such that the product of two $*$-DGAs as given in Eqs.~\eqref{star-d} is itself a $*$-DGA which satisfies Eqs.~\eqref{involprop},~\eqref{dcond}, and~\eqref{dconnew}. This is the graded product which we will employ in Section~\ref{Sec_Prod_New}.

\section{A new product prescription (and its advantages)}
\label{Sec_Prod_New}

In this section we apply the graded tensor product reviewed in Subsection~\ref{Prelim_grad_tens} to redefine the tensor product of two real, spectral triples. We consider the even-even, even-odd, and odd-odd cases separately. Our goal will be to ensure that the product geometries we define always have a well defined KO-dimension. Before we begin it should be noted that in addition to this dimensional requirement, product geometries must also satisfy a number of other geometric conditions in order to qualify as NCGs~\cite{Connes:1996gi,Chamseddine:2006ep,
ConnesMarcolliBook}. We will not discuss these extra  conditions here, but instead refer the reader to the relevant sections of~\cite{2011IJGMM..08.1833D,Cacic:2012qj,Vanhecke99,Sitarz} to see that this will indeed always be the case.

\subsection{The even-even case}
\label{Sec_Prod_ee}

The graded tensor product which we reviewed in Subsection~\ref{Prelim_grad_tens} is directly applicable when constructing a product geometry from two real even spectral triples. For even spectral triples $\{A,H,D,J,\gamma\}$ the Hilbert space $H$ is $\mathbb{Z}_2$ graded, with the degree of its elements distinguished by the grading operator $\gamma$. The degree of the algebra representation $\pi$ with respect to the grading on $H$, and also that of the operators $\{D,J,\gamma\}$ is determined by the NCG axioms, a review of which can be found for example in~\cite{Connes:1996gi,Chamseddine:2006ep,
ConnesMarcolliBook,vanSuijlekomBook,Schucker:2001aa}. The grading operator  is both hermitian and unitary $\gamma = \gamma^* = \gamma^{-1}$, which means that it is equipped with eigenvalues $\pm 1$. We say that elements $h\in H$ which satisfy $\gamma h = h$ are of `even' degree, while elements satisfying $\gamma h = -h$ are of `odd' degree. The representation $\pi$ of the input algebra $A$ on $H$ is even with respect to the grading on $H$, which means that it satisfies $[\pi(a),\gamma]=0$ for all $a\in A$. Meanwhile the Dirac operator is of odd degree with respect to the grading on $H$, which means that it satisfies $\{D,\gamma\}=0$. The degree of the real structure operator depends on the KO-dimension of the geometry:  $J\gamma = \epsilon''\gamma J$. For a more complete discussion of the $\mathbb{Z}_2$ grading on $H$ see also~\cite{Boyle:2016cjt}.

Following the prescription outlined in Subsection~\ref{Prelim_grad_tens}  we define the graded product between two real, even spectral triples $T_i = \{A_i,H_i,D_i,J_i,\gamma_i\}$ and 
$T_j = \{A_j,H_j,D_j,J_j,\gamma_j\}$ as $T_{i,j} = \{A_{i,j},H_{i,j},D_{i,j},J_{i,j},\gamma_{i,j}\}$, where:  
\begin{align}
A_{i,j} = A_i\otimes A_j,\hspace{1.5cm} H_{i,j} = H_i\otimes H_j,\nonumber\\
 D_{i,j} = D_i\otimes \mathbb{I}_j + \mathbb{I}_i\otimes D_j,~~~~~~\label{prodBF}\\
J_{i,j} = J_i\otimes J_j,\hspace{1.5cm}
\gamma_{i,j} = \gamma_i\otimes \gamma_j,~~\nonumber
\end{align}
and where the lack of `hats' indicates that we are using the graded tensor product of Subsection~\ref{Prelim_grad_tens}. We note that the real structure operator in a spectral triple may be viewed as a star operation on the input Hilbert space (as described in~\cite{Boyle:2014wba,Farnsworth:2014vva,
Boyle:2016cjt}), and so the form of the product real structure operator $J_{i,j}$ in~\eqref{prodBF} follows directly from Eq.~\eqref{eq_invol}. Similarly, the Dirac operator of a spectral triple may be understood as deriving from the differential operator of a $*$-DGA (as for example in~\cite{Boyle:2016cjt}), and so the form of $D_{i,j}$ in~\eqref{prodBF} follows directly from Eq.~\eqref{eq_dirac}.

To compare our new definitions with the traditional definitions given in Eqs.~\eqref{Eq_Prod_Orig}, as well as to compare with the product triples defined in~\cite{2011IJGMM..08.1833D,Cacic:2012qj,Vanhecke99,
Sitarz}, we have only to re-express our graded tensor product in terms of the un-graded tensor product, which we do now: Because the representations of the algebras $A_i,A_j$ and grading operators $\gamma_i,\gamma_j$ are of even order, the action of the product algebra $A_{i, j}$ and product grading operator  $\gamma_{i, j}$ given in eq~\eqref{prodBF} may be expressed on $H_{i, j}$ exactly as in eq~\eqref{Eq_Prod_Orig}:
\begin{subequations}
\begin{align}
H_{i, j} = H_i\widehat{\otimes} H_j,\hspace{1.3cm}
A_{i, j} = A_i\widehat{\otimes} A_j,\hspace{1.3cm}
\gamma_{i, j} = \gamma_i\widehat{\otimes} \gamma_j.\label{alggradBF}
\end{align}
The Dirac operators $\{D_i,D_j\}$ however are of odd order, while  the order of the real structure operators $\{J_i,J_j\}$ depends on their KO-dimension signs $\{\epsilon_i'',\epsilon_j'' \}$. Re-expressing the operators $J_{i, j}$ and $D_{i, j}$ of Eqs.~\eqref{prodBF} using the un-graded tensor product results in the appearance of grading operators:
\begin{align}
J_{i,j}= J_i\gamma_i^{(1-\epsilon_j'')/2}\widehat{\otimes} J_j,\hspace{1.3cm}D_{i,j}= D_i\widehat{\otimes} \mathbb{I}_j + \gamma_i\widehat{\otimes} D_j,
\end{align} 
or
\begin{align}
\widetilde{J}_{i,j}= J_i\widehat{\otimes} J_j\gamma_j^{(1-\epsilon_i'')/2},\hspace{1.3cm}
\widetilde{D}_{i,j}= D_i\widehat{\otimes} \gamma_j + \mathbb{I}_i\widehat{\otimes} D_j,
\end{align} \label{defevenprodBF}\end{subequations}
where the two choices depend on the Kozul sign convention chosen (see Subsection~\ref{Prelim_grad_tens}). These two choices are unitarily equivalent, with the unitary transformation given as in eq~\eqref{Eq_transformU}. As would be expected given the unitary equivalence of $\{D_{i, j},J_{i, j}\}$ and $\{\widetilde{D}_{i, j},\widetilde{J}_{i, j}\}$, the signs $\{\epsilon_{i, j},\epsilon_{i, j}',\epsilon_{i, j}''\}$ corresponding to a product triple $T_{i, j}$ do \textit{not} depend on which `Kozul' sign convention is chosen:
\begin{align}
\epsilon_{i, j} = (-1)^{(1-\epsilon_i'')(1-\epsilon_j'')/4}\epsilon_i\epsilon_j,\hspace{1cm}
\epsilon_{i, j}' = \epsilon_i'\epsilon_j''=\epsilon_i''\epsilon_j',\hspace{1cm}
\epsilon_{i, j}'' = \epsilon_i''\epsilon_j''. \label{signsBF}
\end{align}
Comparing with the KO-dimension signs of the traditional product prescription in Eq.~\eqref{Eq_symprod}, the signs in Eq.~\eqref{signsBF} \textit{are} completely symmetric and do not depend on what order the tensor product is taken in (i.e. both $T_{i,j}$ and $T_{j,i}$ are always well defined). Our naming convention for the KO-dimension table now also becomes apparent: The product between two even `upper' (`lower') spectral triples is always well defined and results in an `upper' (`lower') product triple of the correct KO-dimension. One can also check that the product between three `upper' (`lower') spectral triples always remains well defined and is associative. It should be stressed that the graded product automatically organizes the KO-dimension table into a closed set of `upper' and `lower' signs in this way, and this is \textit{not} something we have introduced by hand (i.e. we have \textit{not} made an arbitrary choice such as $\epsilon_L' = +1$ for all even dimensions as is regularly done in the NCG literature). We re-arrange the KO-dimension signs according to our `upper' and `lower' classification in Table~\ref{Tab_KO_comp}, with the `upper' signs for a given KO-dimension placed above the corresponding  `lower' signs. The `upper' signs are those for which $\epsilon_U' = \epsilon_U''$, while the `lower' signs satisfy $\epsilon_L' = -\epsilon_L''$. With this presentation a clear pattern between the signs emerges:   $\{\epsilon_{n+1,U},\epsilon_{n+1,U}'\}=\{\epsilon_{n,L},\epsilon_{n,L}'\}$ (where we remind the reader that for odd KO-dimensions the `upper' and `lower' signs $\{\epsilon,\epsilon'\}$ are degenerate). Every real, even spectral triple is equipped with both an `upper' and a `lower' real structure, and  eqs~\eqref{prodBF} and eqs~\eqref{defevenprodBF} consistently define how to take their product.

\subsection{The even-odd cases}

Our next goal is do define the product between odd and even spectral triples. The Hilbert space $H_{i, j}$, and algebra $A_{i, j}$ will be the same as in Eq.~\eqref{alggradBF}, but now only the even dimensional space will be equipped with a non-trivial grading
operator. We therefore choose $\{D_{i,j},J_{i,j}\}$ or $\{\widetilde{D}_{i,j},\widetilde{J}_{i,j}\}$ from Eqs.~\eqref{defevenprodBF}, according to whether the even triple is the first one or the second one in the product respectively (a similar choice was made in~\cite{2011IJGMM..08.1833D}). Making use of Eqs.~\eqref{defevenprodBF} in this way however presents us with a puzzle: how do we define the $\epsilon''$ signs in odd dimensions? We take inspiration from Clifford algebras\footnote{An irreducible representation of the $d=2n+1$ dimensional Clifford algebra can be constructed by extending the irreducible representation of the $d=2n$ dimensional Clifford algebra by $\gamma$, or alternatively a reducible representation can be constructed as a sub-algebra of a representation of the $d=2n+2$ dimensional Clifford algebra. See~\cite[\S B]{Polchbook} for details.}, and define:
\begin{align}
\epsilon_{n+1,L}'' = \epsilon_{n,U}'',\label{eq_oddsigns}
\end{align}
for all $n\in \mathbb{Z}_8$\footnote{Note that we could have also chosen $\epsilon_{n+1,U}'' = \epsilon_{n,L}''$, which would have resulted in a more aesthetically pleasing presentation of the KO-dimension table, but at the same time would have also propagated various signs through the definition for the tensor product between odd-even and odd-odd spectral triples.}. We have included these additional $\epsilon''$ signs for the odd cases in table~\ref{Tab_KO_comp}. With these definitions in place, the product between an upper (lower) $2n$ dimensional geometry and an upper (lower) $2m+1$ dimensional geometry, according to Eqs.~\eqref{defevenprodBF}, yields a geometry with upper (lower) KO-dimension $2(m+n)+1$.

\begin{table}[h]
\begin{center}
  \begin{tabular}{| c | c | c | c | c | c | c | c | c  |}
    \hline
     & $0$ & $1$ & $2$ & $3$ & $4$ & $5$ & $6$ & $7$ \\ \hline\hline
      $\epsilon$    & $\begin{matrix} +1\\+1\end{matrix}$ & $+1$ & $\begin{matrix} +1\\-1\end{matrix}$ & $-1$ & $\begin{matrix} -1\\-1\end{matrix}$ & $-1$ & $\begin{matrix} -1\\+1\end{matrix}$ & $+1$ \\ 
      \hline
            $\epsilon'$   & $\begin{matrix} +1\\-1\end{matrix}$ & $-1$& $\begin{matrix} -1\\+1\end{matrix}$ & $+1$& $\begin{matrix} +1\\-1\end{matrix}$ & $-1$& $\begin{matrix} -1\\+1\end{matrix}$ & $+1$ \\ 
            \hline
                  $\epsilon''$   & $+1$ & {\color{red}$\begin{matrix} -1\\+1\end{matrix}$} & $-1$ & {\color{red}$\begin{matrix} +1\\-1\end{matrix}$} & $+1$ & {\color{red}$\begin{matrix} -1\\+1\end{matrix}$} & $-1$ &  {\color{red}$\begin{matrix} +1\\-1\end{matrix}$}  \\
                  \hline
  \end{tabular}
\end{center}
\caption{\label{Tab_KO_comp}\textit{Complete Mod 8 KO-dimension table:} Black entries correspond to the KO-dimension signs $\{\epsilon,\epsilon',\epsilon''\}$ of Eq.~\eqref{Eq_Real_Structure}. We introduce the red  $\epsilon''$ entries for odd KO-dimensions  to facilitate the construction of odd-even and odd-odd product geometries.  In this presentation a clear pattern emerges: $\{\epsilon_{n+1,U},\epsilon_{n+1,U}',\epsilon_{n+1,L}''\}=\{\epsilon_{n,L},\epsilon_{n,L}',\epsilon_{n,U}''\}$.} 
\end{table}

Note that the reader may wish to view these new odd $\epsilon''$ signs as corresponding to the two choices $\gamma = \{\mathbb{I},i\mathbb{I}\}$, which leave the upper and lower signs $\{\epsilon,\epsilon'\}$ degenerate and which satisfy $[D,\gamma] =[\gamma,\pi(a)]=0$. While $\gamma = i\mathbb{I}$ no longer satisfies the usual defining condition $\gamma^2 = \mathbb{I}$~\cite{ConnesMarcolliBook}, both choices $\gamma = \{\mathbb{I},i\mathbb{I}\}$ are unitary, which means that we can still make use of the unitary transformation given in Eq.~\eqref{Eq_transformU}\footnote{When $\gamma_j = i\mathbb{I}$ the product Dirac operator $D_{i,j} = D_i\widehat{\otimes}\mathbb{I}_j + \gamma_i\widehat{\otimes} D_j$ transforms as $UD_{i,j}U^* = D_i\gamma_i\widehat{\otimes} i + \gamma\widehat{\otimes} D$, while the real structure operator $J_{i,j} = J\gamma_i^{(1-\epsilon_j)/2}\widehat{\otimes} J_j$ transforms as $UJ_{i,j}U^* = J_i\gamma_i^{(1-\epsilon_i'')/2}\widehat{\otimes} i^{(1-\epsilon_i)/2}J_j$, where $U$ is the unitary given in Eq.~\eqref{Eq_transformU}. When $\gamma_j = \mathbb{I}$ the product Dirac and real structure operators are invariant under the unitary transformation given in Eq.~\eqref{Eq_transformU}: 
$D_{i,j}=UD_{i,j}U^*$, and $J_{i,j} = UJ_{i,j}U^*$.}. In practice however we will never be making any practical use of the identification $\gamma = \{\mathbb{I},i\mathbb{I}\}$ when constructing product geometries (i.e. we will never build a product grading operator $\gamma_{i, j}$ where for example $\gamma_i=i\mathbb{I}$).

\label{Sec_Prod_eo}

\subsection{The odd-odd case}
\label{Sec_Prod_oo}
For the odd-odd cases there is no non-trivial grading operator to work with and so we can no longer make use of the product given in Eq.~\eqref{defevenprodBF}. Taking inspiration from~\cite{2011IJGMM..08.1833D,Cacic:2012qj} however we define: 
\begin{align}
A_{i, j} = A_i\widehat{\otimes} A_j,\hspace{1.5cm} H_{i, j} = H_i\widehat{\otimes} H_j\widehat{\otimes} \mathbb{C}^2,\hspace{1.5cm}
\gamma_{i, j} = \mathbb{I}_i\widehat{\otimes}\mathbb{I}_j\widehat{\otimes} \sigma_3,\hspace{0.7cm}\nonumber\\
D_{i, j} = D_i\widehat{\otimes} \mathbb{I}\widehat{\otimes}\sigma_1 +\mathbb{I}_i\widehat{\otimes} D_j \widehat{\otimes} \sigma_2,\hspace{1.5cm}
J_{i, j} = J_i\widehat{\otimes} J_j\widehat{\otimes} \sigma_1^{(1-\epsilon_i'')/2}(i\sigma_2)^{(1+\epsilon_j'')/2}\circ cc,\label{Eq_Prod_Odd}
\end{align}
where the $\sigma_i$ are Pauli matrices, and once again the signs $\epsilon''$ are determined for odd KO-dimensions using Eq.~\eqref{eq_oddsigns}. The representation of the algebra is understood to be trivial on the $\mathbb{C}^2$ factor, i.e. $\pi(a_i \otimes a_j) =
\pi_i(a_i) \otimes \pi_j(a_j) \otimes \mathbb{I}_{\mathbb{C}^2}$~\cite{2011IJGMM..08.1833D}. With these definitions in place, the product between an upper (lower) $2n+1$ dimensional geometry and an upper (lower) $2m+1$ dimensional geometry yields a geometry with upper (lower) KO-dimension $2(m+n+1)$ without the need for the lookup tables that were required in~\cite{2011IJGMM..08.1833D,Cacic:2012qj}. Finally, just as in the even-even and even-odd cases, the  odd-odd product KO-dimension signs depend symmetrically on their constituent KO-dimension signs:
\begin{align}
\epsilon_{i, j} = (-1)^{(1+\epsilon_i'')(1+\epsilon_j'')/4}\epsilon_i\epsilon_j,\hspace{1.3cm}\epsilon_{i, j}' = -\epsilon_{i}'\epsilon_j'' = -\epsilon_{i}''\epsilon_j',\hspace{1.3cm}\epsilon_{i, j}'' = -\epsilon_i''\epsilon_j''.\label{signsBFodd}
\end{align}

\subsection{A useful Mnemonic}
\label{Prod_Mneminic}
Having defined the product between real spectral triples, we are now able to  introduce a useful mnemonic for `deriving' the full KO-dimension table. We proceed in three steps:

\textbf{Step 1.} There are $2^3=8$ possible sign combinations $\{\epsilon,\epsilon',\epsilon''\}$  corresponding to the even KO-dimension cases, and $2^2=4$ sign combinations $\{\epsilon,\epsilon'\}$ corresponding to the odd KO-dimension cases. Begin by matching the $8$ even cases into $4$ pairs according to the relation $J_U = \gamma J_L$. Note that it is not yet important to know which set of  signs in each pair should be labelled  `upper', and which should be labelled `lower', only which pairs belong together.

\textbf{Step 2.} It is now possible to determine   which of the even sign cases corresponds to KO-dimension $0$ mod $8$, and which of the even sign cases corresponds to KO-dimension $4$ mod $8$. The product of two KO-dimeinsion $0$ mod $8$ spectral triples is again a KO-dimension $0$ mod $8$ spectral triple. This is the only KO-dimension which has this property, and so we can use equations~\eqref{signsBF} directly to distinguish which signs correspond to KO-dimension $0$ mod $8$. Similarly the product of two KO-dimension $4$ mod $8$ spectral triples gives a spectral triple of KO-dimension $0$ mod $8$, and so the signs corresponding to KO-dimension $4$ mod $8$ are also readily distinguishable.

\textbf{Step 3.} Usually modular arithmetic would prevent us from going any further, however the `upper' signs $\{\epsilon_{n+1,U},\epsilon_{n+1,U}'\}$ corresponding to a spectral triple of KO-dimension $n$ mod $8$ match the `lower' signs $\{\epsilon_{n,L},\epsilon_{n,L}'\}$ corresponding to a spectral triple of KO-dimension $n-1$ mod $8$, while these `upper and `lower' signs in the odd cases are degenerate. We therefore have:
\begin{align}
\{\epsilon_{2n,L},\epsilon_{2n,L}'\}=\{\epsilon_{2n+1,U},\epsilon_{2n+1,U}'\}=\{\epsilon_{2n+1,L},\epsilon_{2n+1,L}'\}=\{\epsilon_{2n+2,U},\epsilon_{2n+2,U}'\}.\label{sgnchoiceBF}
\end{align}
Equation~\eqref{sgnchoiceBF} is restrictive enough that it allows the `upper' and `lower' signs of KO-dimension $0$ mod $8$ to be distinguished. Alternatively we could have distinguished `upper' signs from `lower' by noting that for `upper' signs $\epsilon_U' = \epsilon_U''$, while for `lower' signs $\epsilon_L' = -\epsilon_L''$. This is enough information to fill out the remainder of table~\ref{Tab_KO_comp}.

\subsection{Concluding remarks}
\label{Sec_Conclusion}

We conclude this section with a brief recap of the advantages of our graded product of spectral triples, which we introduced in Eqs.~\eqref{prodBF}, \eqref{defevenprodBF} and~\eqref{Eq_Prod_Odd}.
\begin{itemize}
\item \textbf{Well defined products:} The first point to note is that our product is always well defined for any pair of real spectral triples of any KO-dimensions including the odd-odd cases. In particular our product does not rely on the various look-up tables which were required for the odd-odd cases in~\cite{2011IJGMM..08.1833D,Cacic:2012qj}. Furthermore, our product is associative, and as is clear from Eqs.~\eqref{signsBF} and~\eqref{signsBFodd} it is also symmetric in the sense that if a product space $T_{i,j}$ is well defined, then so is $T_{j,i}$.

\item \textbf{Meaning behind the grading factors:} The authors in~\cite{Sitarz,Vanhecke99,Cacic:2012qj} all found ways of cleverly inserting grading factors into their definitions for the product of real structure operators in order to construct well defined product spectral triples. In our formulation the appearance of grading factors in both the Dirac  and real structure operators is natural, and is no longer a mystery. They  result automatically when translating between the graded tensor product and the ungraded tensor product.

\item \textbf{Transformation under unitaries:} Our product always remains well defined under the unitary transformation given in Eq.~\eqref{Eq_transformU}. Unlike in previous work, we stress that the Dirac operator \textit{and} the real structure operator of an even spectral triple transform non-trivially under the action of the unitary operator given in Eq.~\eqref{Eq_transformU}. In addition the  unitary  equivalence of the two choices $\{D_{i,j},J_{i,j}\}$ and $\{\widetilde{D}_{i,j},\widetilde{J}_{i,j}\}$ is linked to the choice of Kozul sign in the graded tensor product.

\item \textbf{KO-Dimension patterns:} Our product naturally distinguishes the `upper' from the `lower' KO-dimension signs. What is more, once this naming convention is adopted a number of patters emerge in the KO-dimension table which were previously obscured by the arbitrary distinction between KO-dimension signs for which $\epsilon'=+1$ and those for which $\epsilon' = -1$.

\item\textbf{KO-dimension table extension:} Our product suggests a natural extension of the KO-dimension table, in which there are $8$ rather than $4$ possible odd KO-dimension combinations.

\end{itemize}

We close with the product table corresponding to our prescription in Table~\ref{evenevenbarD}. In order to appreciate just how simple our product is, comparison should be made for example with Tables 2-5 of~\cite{2011IJGMM..08.1833D} and tables 2-5 of~\cite{Vanhecke99}. Our product for the odd-odd cases also avoids the need for lookup tables which can be seen for example in Table 6 of~\cite{2011IJGMM..08.1833D}, and Table 2.3 of~\cite{Cacic:2012qj}.

\begin{table}[h!]
\centering
\begin{tabular}{|l||*{8}{c|}|*{8}{c|}}\hline
 &$0_U$&$1_U$&$2_U$
&$3_U$&$4_U$&$5_U$&$6_U$
&$7_U$
&$0_L$&$1_L$&$2_L$
&$3_L$&$4_L$&$5_L$&$6_L$
&$7_L$
\\\hline\hline
$0_U$ &$0_U$&$1_U$&$2_U$
&$3_U$& $4_U$ & $5_U$ & $6_U$ 
&$7_U$& & 
& & & & 
&  & \\\hline
$1_U$ &$1_U$&$2_U$&$3_U$
&$4_U$& $5_U$ & $6_U$ & $7_U$ 
&$0_U$& & 
& & & & 
&  &\\\hline
$2_U$ &$2_U$&$3_U$&$4_U$
&$5_U$& $6_U$ & $7_U$ & $0_U$ 
&$1_U$& & 
& & & & 
&  &\\\hline
$3_U$ &$3_U$&$4_U$&$5_U$
&$6_U$& $7_U$ & $0_U$ & $1_U$ 
&$2_U$& & 
& & & & 
&  &\\\hline
$4_U$ &$4_U$&$5_U$&$6_U$
&$7_U$& $0_U$ & $1_U$ & $2_U$ 
&$3_U$& & 
& & & & 
&  &\\\hline
$5_U$ &$5_U$&$6_U$&$7_U$
&$0_U$& $1_U$ & $2_U$ & $3_U$ 
&$4_U$& & 
& & & & 
&  &\\\hline
$6_U$ &$6_U$&$7_U$&$0_U$
&$1_U$& $2_U$ & $3_U$ & $4_U$ 
&$5_U$& & 
& & & & 
&  &\\\hline
$7_U$ &$7_U$&$0_U$&$1_U$
&$2_U$& $3_U$ & $4_U$ & $5_U$ 
&$6_U$& & 
& & & & 
&  &

\\\hline\hline
$0_L$ & & & & & & & & & $0_L$  & $1_L$ & $2_L$ 
& $3_L$ & $4_L$ & $5_L$ & $6_L$ 
& $7_L$ \\\hline
$1_L$ & & & & & & & & & $1_L$  & $2_L$ & $3_L$ 
& $4_L$ & $5_L$ & $6_L$ & $7_L$ 
& $0_L$ \\\hline
$2_L$ & & & & & & & & & $2_L$  & $3_L$ & $4_L$ 
& $5_L$ & $6_L$ & $7_L$ & $0_L$ 
& $1_L$ \\\hline
$3_L$ & & & & & & & & & $3_L$  & $4_L$ & $5_L$ 
& $6_L$ & $7_L$ & $0_L$ & $1_L$ 
& $2_L$ \\\hline
$4_L$ & & & & & & & & & $4_L$  & $5_L$ & $6_L$ 
& $7_L$ & $0_L$ & $1_L$ & $2_L$ 
& $3_L$
\\\hline
$5_L$ & & & & & & & & & $5_L$ & $6_L$ & $7_L$ 
& $0_L$ & $1_L$ & $2_L$ & $3_L$ 
& $4_L$ \\\hline
$6_L$ & & & & & & & & & $6_L$ & $7_L$ & $0_L$ 
& $1_L$ & $2_L$ & $3_L$ & $4_L$ 
& $5_L$ \\\hline
$7_L$ & & & & & & & & & $7_L$ & $0_L$ & $1_L$ 
& $2_L$ & $3_L$ & $4_L$ & $5_L$ 
& $6_L$ \\\hline
\end{tabular}
\caption{The graded product table for real spectral triples. }
\label{evenevenbarD}
\end{table}

Note: During the write-up of this work we learned that the authors C.~Brouder, N.~Bizi and F.~Besnard have also constructed a product of spectral triples similar to that of~\cite{Vanhecke99,Sitarz} for  Lorentzian spectral triples, which they will likely publish along with future work. We make note of this as their work has some similarities to our own which were obtained independently.

Acknowledgements: We would like to thank John Barrett, Nadir Bizi, Latham Boyle,  Christian Brouder, and
Matilde Marcolli for useful discussions during the
writing of this work. This work was supported by the Max Planck Society, and in part by the European Cooperation in Science and Technology association.


\end{document}